\documentclass[12pt,english]{article}
\usepackage{epsf}
\usepackage{geometry}
\usepackage{setspace}
\usepackage{amssymb}

\makeatletter

 \newcommand{\lyxaddress}[1]{
   \par {\raggedright #1
   \vspace{1.4em}
   \noindent\par}
 }

\usepackage{setspace}

\makeatother
\begin{document}

\title{\textbf{\large Reversals in nature and the nature of reversals}}

\author{F. STEFANI$^{\star}\dagger$, M. XU$\dagger$, L. SORRISO-VALVO$\ddagger$, G. GERBETH$\dagger$,\\ and U. G\"UNTHER$\dagger$}

\maketitle

\lyxaddress{\begin{center}$\dagger${\small Forschungszentrum Dresden-Rossendorf, P.O. Box 510119,
D-01314 Dresden,
Germany}\\
$\ddagger${\small LICRYL - INFM/CNR, Ponte P. Bucci, Cubo 31C, 87036 Rende (CS), Italy}
\end{center}}

{\footnotesize
The asymmetric shape of reversals of the Earth's magnetic field
indicates
a possible connection with relaxation oscillations as they were
early discussed by van der Pol.
A simple mean-field dynamo model with a spherically symmetric $\alpha$
coefficient is analysed with view on this similarity, and a comparison
of the time series and the phase space trajectories with those of
paleomagnetic measurements is carried out. For highly supercritical
dynamos a
very good agreement with the data is achieved. Deviations of
numerical reversal
sequences from Poisson statistics are analysed and compared with
paleomagnetic data. The role of the inner core
is discussed in a spectral theoretical context and arguments
and numerical evidence is compiled that the growth of the inner
core might be important for the long term changes of the
reversal rate and the
occurrence of superchrons.}

\noindent \textit{\footnotesize Keywords:} {\footnotesize Dynamo;
$\alpha$ effect; Earth's magnetic field reversals; inner core; 
Poisson statistics}{\footnotesize \par}

\section{Introduction}

It is well known from paleomagnetic measurements that the
axial dipole component of the Earth's magnetic
field has reversed its polarity many times (Merrill \textit{et al.} 1996).
The last
reversal occurred approximately
780000 years ago. Averaged over the last few million years the
mean rate of reversals is approximately 5 per Myr.
At least two superchrons have been identified
as periods of some tens of million years containing
no reversal at all. These are the Cretaceous
superchron extending from approximately 118 to 83 Myr, and the
Kiaman superchron extending approximately from
312 to 262 Myr. Only recently,
the existence of a third superchron has been hypothesized for the
time period between 480-460 Myr (Pavlov and Gallet 2005).

Apart from their irregular occurrence,  one
of the most intriguing features of reversals is
the  pronounced asymmetry in the sense that the
decay of the dipole is much slower than the subsequent
recreation of the dipole with opposite
polarity (Valet and Meynadier 1993, Valet \textit{et al.} 2005).
Observational data also indicate a possible correlation of the
field intensity with the interval between subsequent
reversals (Cox 1968, Tarduno \textit{et al.} 2001, Valet \textit{et al.} 2005).
A further hypothesis concerns the so-called bimodal distribution
of the Earth's virtual axial dipole moment (VADM) with two peaks
at approximately 4 $\times$ 10$^{22}$ Am$^2$
and at  twice that value
(Perrin and Shcherbakov 1997, Shcherbakov \textit{et al.} 2002,
Heller \textit{et al.} 2003).
There is an ongoing discussion about preferred
paths of the virtual geomagnetic pole (VGP)
(Gubbins and Love 1998), and about
differences of the
apparent duration of reversals when seen from sites at
different latitudes
(Clement 2004).

The reality of reversals is quite complex and there
is little hope to understand all their details within a simple
model. Of course, computer simulations of the geodynamo in
general and
of reversals in particular (Wicht and Olson 2004,
Takahashi \textit{et al.} 2005)
have progressed much since the first
fully coupled  3D simulations of a reversal  by Glatzmaier and
Roberts in 1995.
However, the severe problem
that simulations have to work
in parameter regions (in particular
for the magnetic Prandlt number and the Ekman number)
which are far away from the real values of the Earth,
will remain for a long time.

With view on these problems to carry out
realistic geodynamo simulations,
but also with a side view on the recent
successful dynamo experiments
(Gailitis \textit{et al.} 2002), it is
legitimate to ask what are
the most essential ingredients for a dynamo to
undergo reversals in a similar way as the
geodynamo does.

Roughly speaking, there are two classes of simplified models
which
try to explain reversals (Fig. 1). The first one relies on the
assumption that the
dipole field is somehow rotated from a given orientation to the
reversed one via
some intermediate state (or states).
The rationale
behind this ``rotation model''
is that many kinematic dynamo simulations
have revealed an approximate degeneration of the axial and the
equatorial dipole (Gubbins \textit{et al.} 2000,
Aubert \textit{et al.} 2004). A similar
degeneration
is also responsible for the
appearance of hemispherical dynamos in dynamically coupled models
(Grote and Busse 2000). In this case, both quadrupolar and dipolar
components contribute nearly equal magnetic energy so that their
contributions
cancel in one hemisphere and add to each other in the
opposite hemisphere.

The interplay between the nearly degenerated
axial dipole, equatorial dipole, and quadrupole was used in
a simple model to explain reversals and
excursion in one
common scheme
(Melbourne \textit{et al.} 2001). It is interesting that
this model of non-linear interaction of the three modes provides
excursion and reversals without the inclusion of noise.

A second  model belonging to this class
is the model developed by  Hoyng and coworkers
(Hoyng \textit{et al.} 2001, Schmidt \textit{et al.} 2001,
Hoyng and Duistermaat 2004)
which deals with one steady axial dipole mode coupled via noise
to an oscillatory ``overtone'' mode. Again, the idea
is that the magnetic energy is taken over in an
intermediate time by this
additional mode.

\begin{figure}
\begin{center}
\epsfxsize=12cm
\epsffile{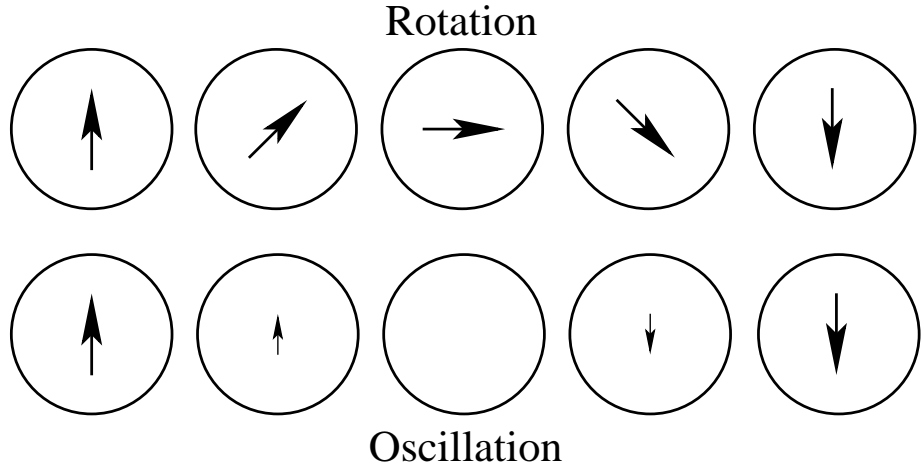}
\end{center}
\caption{Two different principles to explain
reversals in simple models.}
\end{figure}

Another tradition to explain reversals relies
on the very specific interplay between a steady and
an oscillatory branch
of the dominant axial dipole mode.
This idea was expressed by Yoshimura et al. (1984)
and by a number of other authors
(Weisshaar 1982, Sarson and Jones 1999,
Phillips 1993, Gubbins and Gibbons 2002).

In  a series of recent papers we have tried to exemplify this
scenario within a strongly simplified mean-field dynamo model.
The starting point was the observation
(Stefani and Gerbeth 2003)
that even simple $\alpha^2$ dynamos with a spherically
symmetric helical turbulence parameter $\alpha$
can have oscillatory dipole solutions, at least
for a certain variety of profiles $\alpha(r)$ characterized
by a sign change along the radius.
This restriction to spherically symmetric $\alpha$ models allows to
decouple the axial dipole mode from all other modes which reduces
drastically
the numerical effort to solve the induction equation.
Hence, the general  idea that reversals have to
do with transitions between the steady and the
oscillatory branch {\it of the same eigenmode}
could be studied in great detail and with long time series.

In (Stefani and Gerbeth 2005) it was shown that such a
spherically symmetric dynamo model, complemented by a simple
saturation model and subjected to noise,
can exhibit some of the above mentioned reversal features.
In particular, the model produced asymmetric
reversals, a positive correlation of field strength and
interval length, and a bimodal field
distribution. All these features were attributable
to the magnetic field
dynamics in the vicinity of a branching point (or
{\it exceptional point}) of the spectrum
of the non-selfadjoint dynamo operator where two real
eigenvalues
coalesce and continue as a complex conjugated pair of
eigenvalues. Usually, this exceptional point is associated with
a nearby local maximum of the
growth rate situated at a slightly lower magnetic Reynolds number.
It is essential that  the negative slope of the growth rate
curve between this local
maximum and the exceptional point makes even stationary
dynamos vulnerable to  noise.
Then, the instantaneous
eigenvalue is driven
towards the exceptional point and beyond into the
oscillatory branch where the sign change happens.

An evident weakness of this reversal model
in the slightly supercritical regime was the
apparent necessity to fine-tune the
magnetic Reynolds number and/or the radial profile $\alpha(r)$
in order to adjust the dynamo operator spectrum in an
appropriate way.
However, in a follow-up paper (Stefani  {\textit et al.} 2006a),
it was shown that
this artificial fine-tuning becomes superfluous in the
case of higher supercriticality of the dynamo
(a similar effect was already found for disk dynamos
by  Meinel and Brandenburg (1990)).
Numerical examples and physical arguments
were compiled to show that, with increasing magnetic
Reynolds number,
there is a strong tendency for the exceptional point and the
associated local maximum to
move close  to the zero growth rate line where the
indicated reversal scenario can be actualized.
Although exemplified again
by the spherically symmetric $\alpha^2$ dynamo model,
the main idea of this ``self-tuning'' mechanism
of saturated dynamos into a reversal-prone state
seems well transferable to other dynamos.

In another paper (Stefani {\textit et al.} 2006b) we have
compared
time series of the dipole moment
as they result from our simple model with the
recently published time series
of the last five reversals which occurred during the last 2 Myr
(Valet {\textit et al.} 2005).
Both  the relevant time scales and the typical asymmetric
shape of the paleomagnetic reversal sequences were
reproduced within our simple model. Note that this
was achieved strictly on the basis
of the molecular resistivity of the core material,
without taking resort to any turbulent resistivity.

An important ingredient of this sort of reversal models, the
sign
change of $\alpha$ along the radius, was also utilized
in the reversal model of Giesecke et al. (2005a).
This model is much closer to reality since it includes
the usual North-South asymmetry of $\alpha$.
Interestingly, the sign change of $\alpha$ along the radius
is by no means unphysical but results indeed from simulations
of magneto-convection which were carried out by the same
authors (Giesecke \textit{et al.} 2005b).

It is one of the goals of the present paper to make a
strong case for the ``oscillation model'' in contrast to the
''rotation model''. We will show that the
asymmetry of the reversals is very well recovered by our
model in the case of high (although not very high)
supercriticality.
At the same time, we will show that the ``rotation model''
in the version presented by Hoyng and Duistermaat (2004) leads
to the wrong asymmetry of reversals.

We will start our argumentation by elucidating
the intimate connection of reversals  with  relaxation
oscillations
as they are well known from the van der Pol oscillator
(van der Pol 1926).
Actually, the connection of dynamo solutions with
relaxation oscillations had been
discussed in the context of the 22 year cycle  of
Sun's magnetic field
(Mininni  \textit{et al.} 2000., Pontieri \textit{et al.} 2003)
and its similarity with reversals was brought to our
attention by Clement Narteau (2005).
We will provide evidence for a strong similarity of the dynamics
of the van der Pol oscillator and of our simple dynamo model.
We will also reveal the same  relaxation oscillation character
in the above mentioned reversals data.

In a second part we will discuss some statistical properties
of our reversal model within the context of real reversal data.
We will focus on the ``clustering property'' of reversals
which was observed by Carbone  \textit{et al.} (2006) and further
analyzed by Sorriso-Valvo  \textit{et al.} (2006).

A third focus of this paper will lay on the influence
of the inner core on the reversal model.
One of the usually adopted hypotheses on the role
of the inner core goes back to  Hollerbach and Jones
(1993 and 1995)
who claimed the conductivity of the inner core
might prevent the magnetic field
from more frequent reversals.
We will discuss  the core influence in
terms of a particular spectral
resonance phenomenon that was observed
by G\"unther and Kirillov (2006).
As a consequence, an increasing inner core can even
favor reversals before it starts to
produce more
excursions than reversals.
This will bring us finally to a speculation on the
determining role the inner core growth might have on
the long term reversal rate and on the (quasi-periodic?)
occurrence of superchrons.

\section{Reversals and relaxation oscillations}

In this section, we will characterize the time
series of simple mean field dynamos as a typical
example of
relaxation oscillations.
In addition, we will delineate an alternative model
which was proposed by Hoyng and Duistermaat (2004)
as a toy model of reversals.
The connection with real reversal sequences will be discussed 
in the last subsection.

\subsection{The models}

\subsubsection{Van der Pol oscillator}
The van der Pol equation is an ordinary differential equation
of second order which describes self-sustaining oscillations.
This equation was shown to arise in circuits containing vacuum
tubes (van der Pol 1926) and relies on the fact that
energy is fed into small oscillations and removed from large
oscillations. It is given
by
\begin{eqnarray}
\frac{\partial^2 y}{\partial t^2}&=&\mu(1-y^2)\frac{\partial y}{\partial t}-y
\end{eqnarray}
which  can be rewritten into an equivalent system of two
coupled first order differential equations:
\begin{eqnarray}
\frac{\partial y}{\partial t}&=&z\\
\frac{\partial z}{\partial t}&=&\mu(1-y^2)z-y \; .
\end{eqnarray}
Evidently, the case $\mu=0$ yields harmonic oscillations 
while increasing $\mu$ provides increasing
anharmonicity.

\subsubsection{Spherically symmetric $\alpha^2$ dynamo model}

For a better understanding of the reversal process we have decided 
to use a toy model which is simple  enough to allow for simulations of 
very long time series (in order to do  reasonable statistics),
but which is also capable to capture distinctive  features of
hydromagnetic dynamos, in particular the typical non-trivial saturation 
mechanism via deformations of the dynamo source.
Both requirements together are fulfilled by a 
mean-field dynamo model of $\alpha^2$ type
with a supposed
spherically symmetric, isotropic helical turbulence 
parameter $\alpha$ (Krause and R\"adler 1980). Of course, we are well 
aware of the fact that a reasonable simulation of the Earth's dynamo
should at least account for the North-South asymmetry of $\alpha$
which is not respected in our model. Only in the last but
one section we will discuss certain spectral features of such a
more realistic model.

Starting from the induction equation for the magnetic
field $\bf B$
\begin{eqnarray}
\frac{\partial \bf{B}}{\partial t} ={\bf \nabla \times (\alpha {\bf{B}})} +
(\mu_0 \sigma)^{-1} \nabla^2 {\bf{B}} \; ,
\end{eqnarray}
(with magnetic permeability
$\mu_0$ and electrical conductivity $\sigma$) 
we decompose ${\bf{B}}$ into a
poloidal and a toroidal component, according to
${\bf{B}}=-\nabla \times ({\bf{r}} \times
\nabla S)-{\bf{r}} \times
\nabla T $. Then we expand the defining   scalars $S$ and $T$
in spherical harmonics of degree $l$ and order $m$
with expansion coefficients
$s_{l,m}(r,t)$ and $t_{l,m}(r,t)$.
As long as we remain within the framework of 
a spherically symmetric and isotropic $\alpha^2$ dynamo
problem,
the induction equation
decouples for each degree $l$ and order $m$ into the following pair
of equations
\begin{eqnarray}
\frac{\partial s_l}{\partial \tau}&=&
\frac{1}{r}\frac{\partial^2}{\partial r^2}(r s_l)-\frac{l(l+1)}{r^2} s_l
+\alpha(r,\tau) t_l \; ,\\
\frac{\partial t_l}{\partial \tau}&=&
\frac{1}{r}\frac{\partial}{\partial r}\left( \frac{\partial}{\partial r}(r t_l)-\alpha(r,\tau)
\frac{\partial}{\partial r}(r s_l) \right)
-\frac{l(l+1)}{r^2}
[t_l-\alpha(r,\tau)
s_l] \; .
\end{eqnarray}
Since these equations are independent of the order $m$, we
have skipped $m$ in the index of $s$ and $t$.
The boundary conditions are
$\partial s_l/\partial r |_{r=1}+{(l+1)} s_l(1)=t_l(1)=0$.
In the following we consider only the dipole field with $l=1$.

We will focus on kinematic radial profiles
$\alpha_{kin}(r)$ with a  sign change along the radius which
had been shown to exhibit oscillatory behaviour (Stefani and Gerbeth 2003).
Saturation is enforced by assuming the kinematic
profile $\alpha_{kin}(r)$ to be algebraically
quenched
by the  magnetic field energy averaged over the angles which
can be expressed in terms of $s_{l}(r,\tau)$ and $t_l(r,\tau)$.
Note that this averaging over the angles represents
a severe simplification
since in reality
(even for an assumed spherically
symmetric kinematic $\alpha$) the energy dependent
quenching would result in a breaking of the spherical symmetry.

In addition to this quenching, the $\alpha(r)$ profiles are
perturbed  by
"blobs" of noise which are
considered constant within a correlation
time $\tau_{corr}$.
Physically, such a  noise term  can
be understood
as a consequence of changing boundary
conditions for the flow in the outer core, but also as a substitute
for the omitted influence of
higher multipole modes on the dominant axial dipole mode.

In summary, the $\alpha(r,\tau)$ profile  entering Eqs. (5) and (6)
is written as
\begin{eqnarray}
\alpha(r,\tau)&=&\frac{\alpha_{kin}(r)}{1+
E^{-1}_{mag,0} \left[ \frac{2 s_1^2(r,\tau)}{r^2}+
\frac{1}{r^2}\left( \frac{\partial (r s_1(r,\tau))}
{\partial r} \right)^2
+t_1^2(r,\tau) \right] }     \nonumber\\
&& +\xi_1(\tau) +\xi_2(\tau) \; r^2 +\xi_3(\tau) \; r^3+\xi_4(\tau) \; r^4 \; ,
\end{eqnarray}
where the noise correlation is given by
$\langle \xi_i(\tau) \xi_j(\tau+\tau_1)
\rangle = D^2 (1-|\tau_1|/\tau_{corr}) \Theta(1-|\tau_1|/\tau_{corr})
\delta_{ij}$.
In the following, $C$ will characterize the amplitude of $\alpha$,
$D$ is the noise strength,
and $E_{mag,0}$ is a constant measuring the mean
magnetic field energy.

\subsubsection{An alternative: The model of Hoyng and Duistermaat}

Later we will compare our results with the
results of the model
introduced by Hoyng and Duistermaat (2004)
which describes a steady
axial dipole ($x$)  coupled by multiplicative noise to a
periodic ``overtone'' ($y+iz$). It is given by the ordinary 
differential equation system:
\begin{eqnarray}
\partial_t x&=&(1-x^2) x+V_{11} x+V_{12} y+ V_{13} z\\
\partial_t y&=&-ay-cz+V_{21} x+V_{22} y+ V_{23} z\\
\partial_t z&=&cy-az+V_{31} x+V_{32} y+ V_{33} z\; .
\end{eqnarray}
Evidently, without noise terms $V_{ik}$ this systems decouples into
a steady axial dipole and a periodic ``overtone''.

\subsection{Numerical results in the noise-free case}
In Fig. 2 we show the numerical solutions of
the van der Pol oscillator for different values of $\mu$,
compared with the solutions
of our dynamo model with a kinematic $\alpha$ profile according to
$\alpha_{kin}(r)=1.916 \cdot C  \cdot  (1-6 \; r^2+5 \; r^4)$
(the factor 1.916 results simply from
normalizing the radial
average of $|\alpha(r)|$ to the corresponding
value for constant $\alpha$). In the dynamo model we
vary the dynamo number $C$ in the noise-free case (i.e. $D=0$).
As already noticed, in the van der Pol oscillator
we get
a purely harmonic oscillation for $\mu=0$  which becomes
more and more
anharmonic for
increasing $\mu$.  A quite
similar behaviour can be observed for the
dynamo model.
At $C=6.8$ (which is only slightly above the critical value
$C_c=6.78$) we
obtain a nearly
harmonic oscillation which becomes also more and more
anharmonic for increasing $C$.
The difference to the van der Pol
system is that  at a  certain value of $C$ the
oscillation stops at all and is replaced  by a steady solution ($C=7.24$).
\begin{figure}
\begin{tabular}{cc}
\epsfxsize=7cm
\epsffile{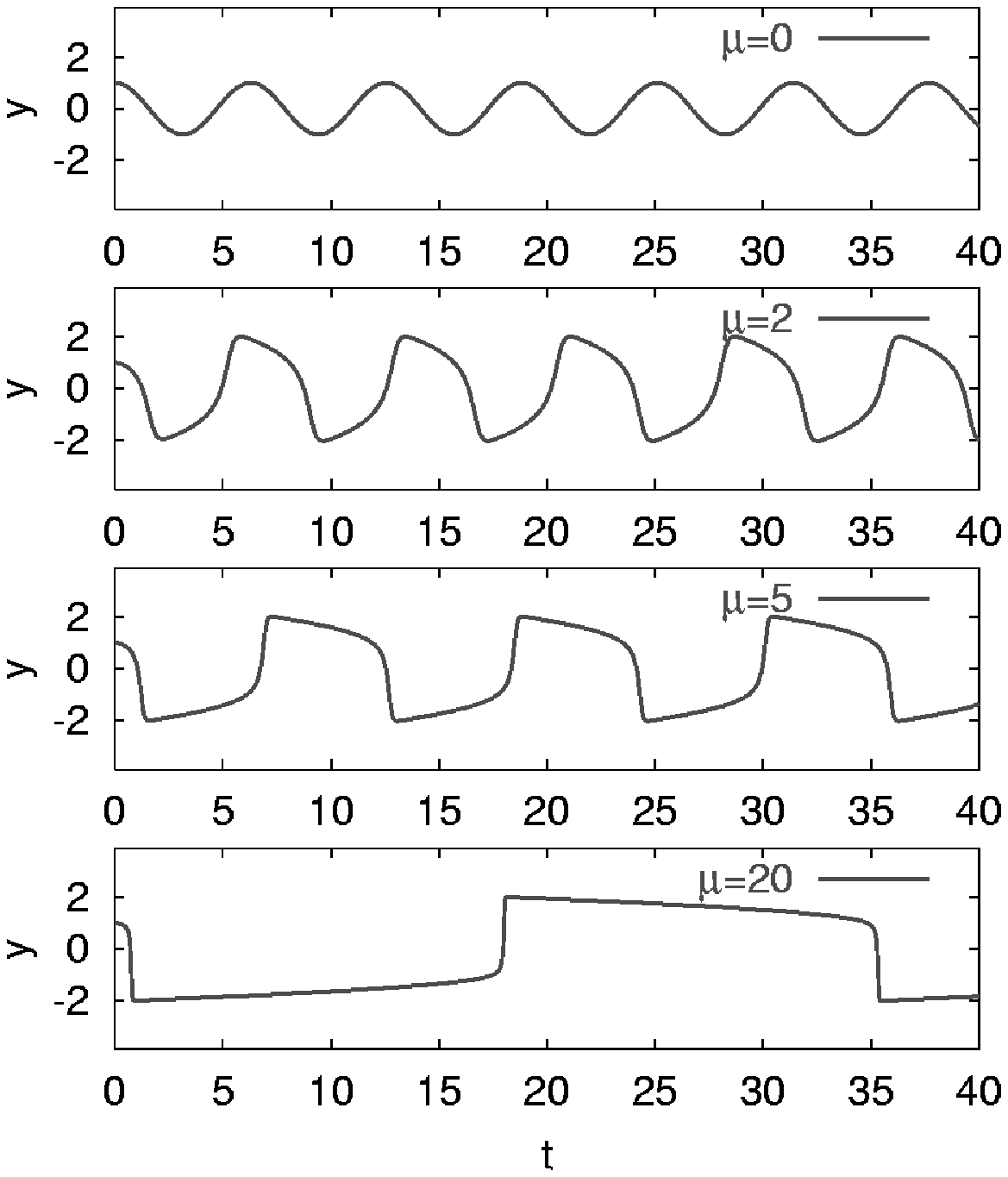}&
\epsfxsize=7cm
\epsffile{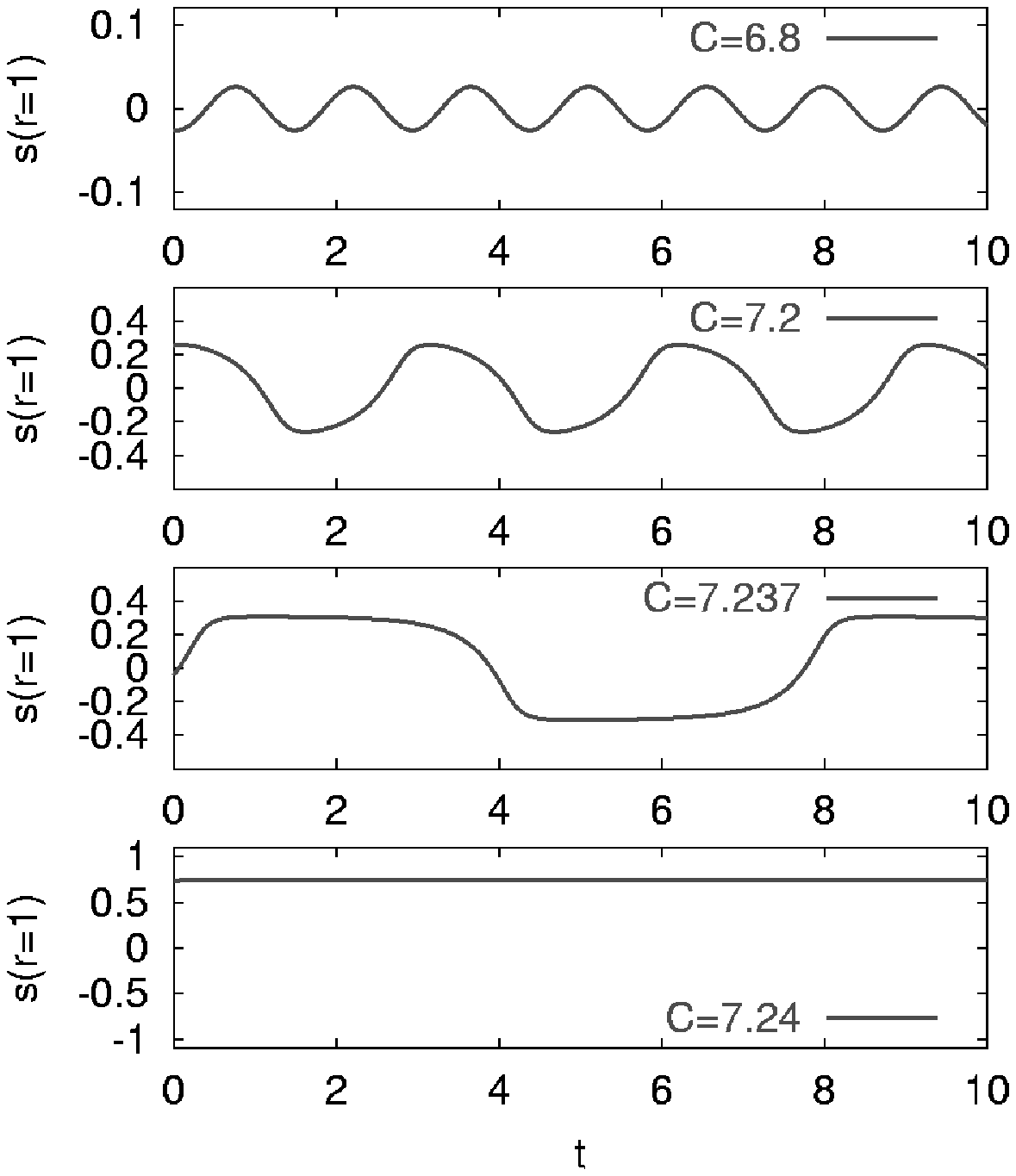}
\end{tabular}
\caption{Typical relaxation oscillations in the van der Pol
model (left) and in the $\alpha^2$ dynamo model (right). With increasing
$\mu$ and $C$, respectively, one observes an increasing
degree of anharmonicity. In the dynamo case, this even goes over
into a steady state.}
\end{figure}

For the two  values $\mu=2$ and $C=7.237$, respectively, we
analyze the time evolution in more detail in Fig. 3. The upper
two panels show $y(t)$ and $y'(t)=z(t)$ in
the van der Pol case and $s(r=1)$ and the time derivative
$s'(r=1)$ in the dynamo case.
The lower two panels show the real and
imaginary parts of the instantaneous eigenvalue $\lambda(t)$.
A note is due here on the definition and the usefulness
of such instantaneous
eigenvalues. For the dynamo problem they result
simply from inserting the instantaneous
quenched $\alpha(r,t_i)$ profiles at the instant $t_i$
into the
time-independent eigenvalue equation. Quite formally,
we could do the same for the van der Pol system by
replacing in the nonlinear term in Eq. (3) $y^2$ by $y^2(t_i)$.
Of course, for a non-linear dynamical system it is
more significant to consider the eigenvalues of the
instantaneous Jacobi matrix\footnote{We recall that the Jacobi matrix
${\bf J}(\bf{x}_0)$ is given by the first-order derivative
terms in a Taylor expansion of a nonlinear vector valued function
${\bf F}({\bf x})={\bf F}({\bf x}_0)+{\bf J}({\bf x}_0)({\bf x}-{\bf x}_0)+O([{\bf x}-{\bf x}_0]^2)$
and characterizes a local linearization
in the vicinity of a given ${\bf x}_0$ (see Childress and Gilbert 1995).} which reads
\begin{eqnarray}
{\bf J}(y(t_i),z(t_i))=\left[ \begin{array}{cc} 0&1\\-1-
2\mu y(t_i) z(t_i)& \mu(1-y^2(t_i))\end{array}        \right]
\end{eqnarray}
In the third and fourth line of the left panel of Fig.
3 we show the real and imaginary parts
of this instantaneous Jacobi matrix eigenvalues,
together with the
eigenvalues resulting from the formal replacement
of $y^2$ by $y^2(t_i)$. It is not surprising that the latter
shows a closer similarity to the
instantaneous eigenvalues for the
dynamo problem which are exhibited in the right panel.

Despite the slight differences,  we see in any case
that during the reversal
there appears a certain interval
characterized by a complex instantaneous eigenvalue
which is ``born'' at an exceptional point
where two real eigenvalues
coalesce and which splits off again at a second
exceptional point into two real eigenvalues.
\begin{figure}
\begin{tabular}{cc}
\epsfxsize=7cm
\epsffile{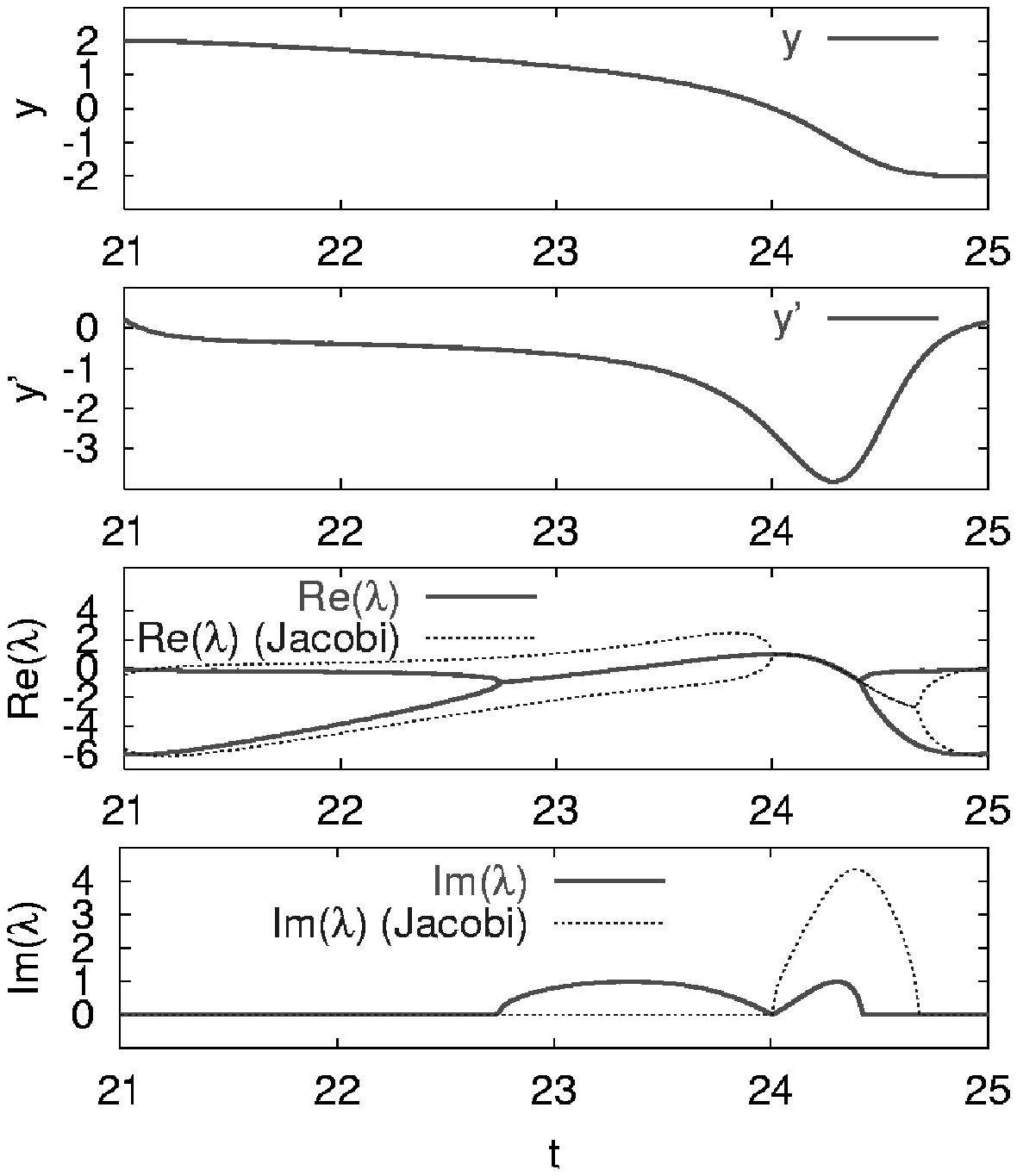}&
\epsfxsize=7cm
\epsffile{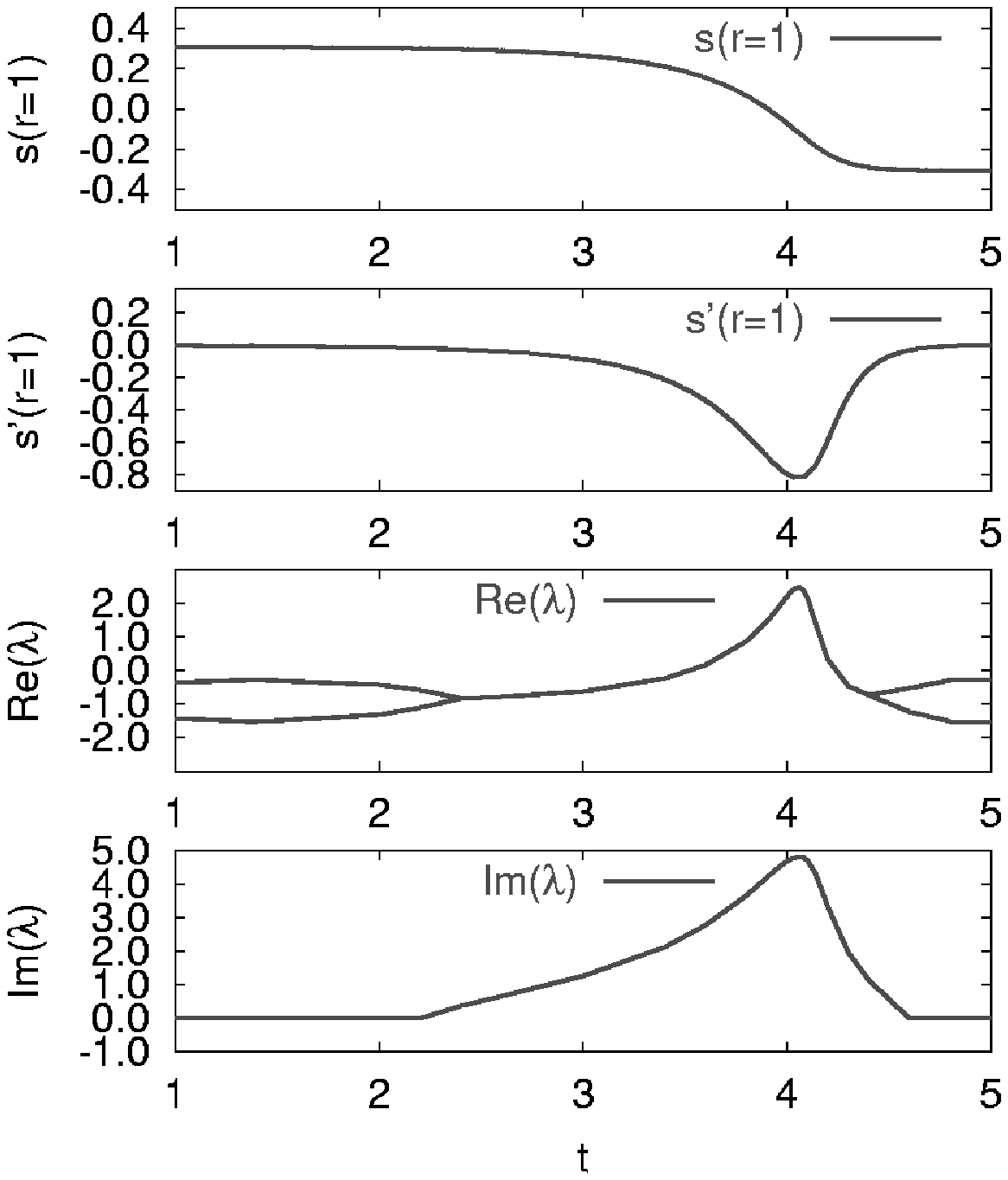}
\end{tabular}
\caption{Details and interpretation of the special cases
$\mu=5$ and $C=7.237$ from Fig. 2, respectively.
From top to bottom the panels show the main signal $y$ and
$s(r=1)$, their time derivatives, and
the real and imaginary parts of the
instantaneous eigenvalues.
In the van der Pol case, the latter are given in the correct version of
the eigenvalue of the Jacobi matrix, and in the formal way which is also
shown for the dynamo case. }
\end{figure}

It is also instructive to show the trajectories in the
phase space, both for the van der Pol and for the
dynamo problem. Figure 4 indicates that the systems are quite
similar.
\begin{figure}
\begin{tabular}{cc}
\epsfxsize=7cm
\epsffile{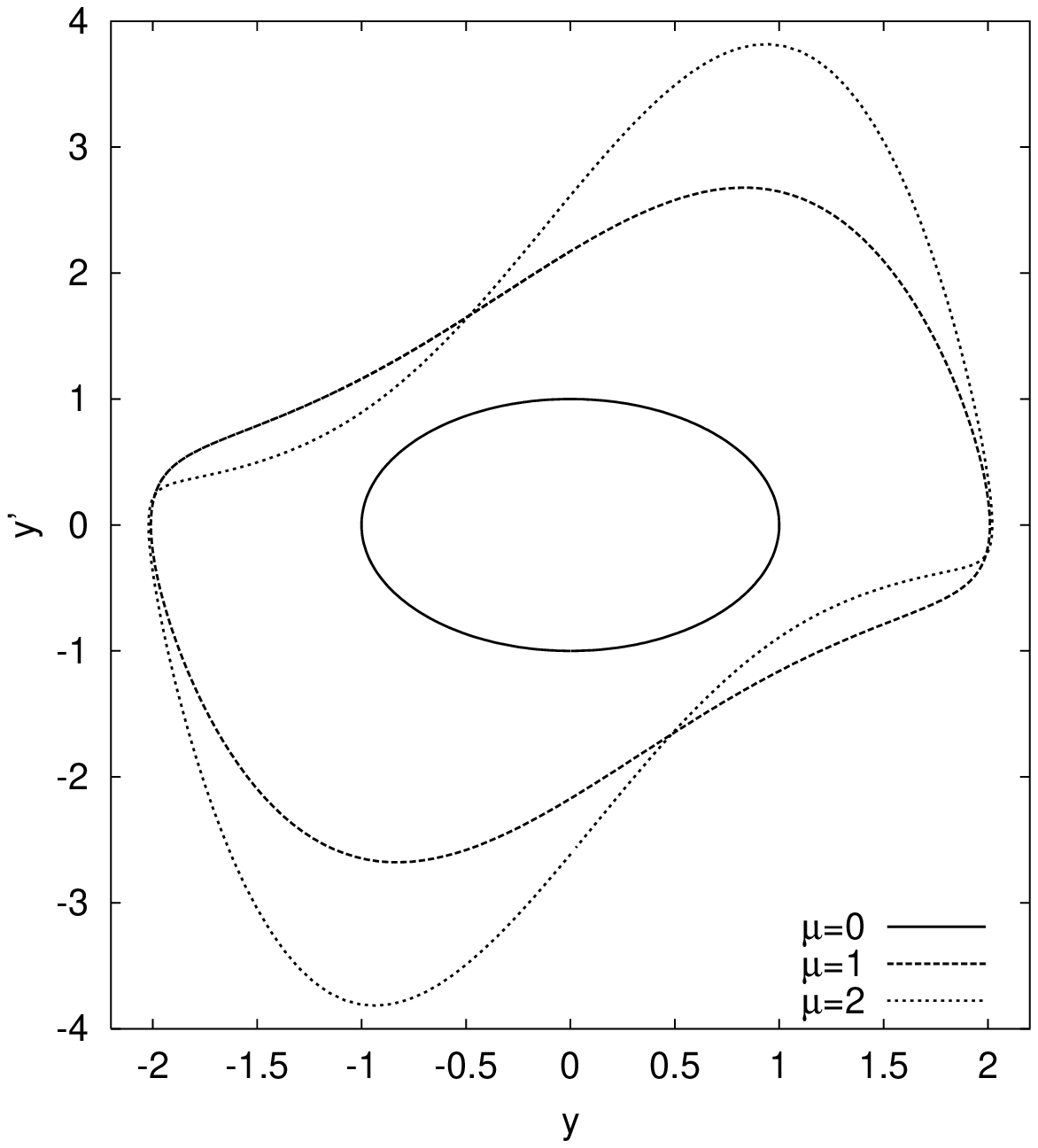}&
\epsfxsize=7cm
\epsffile{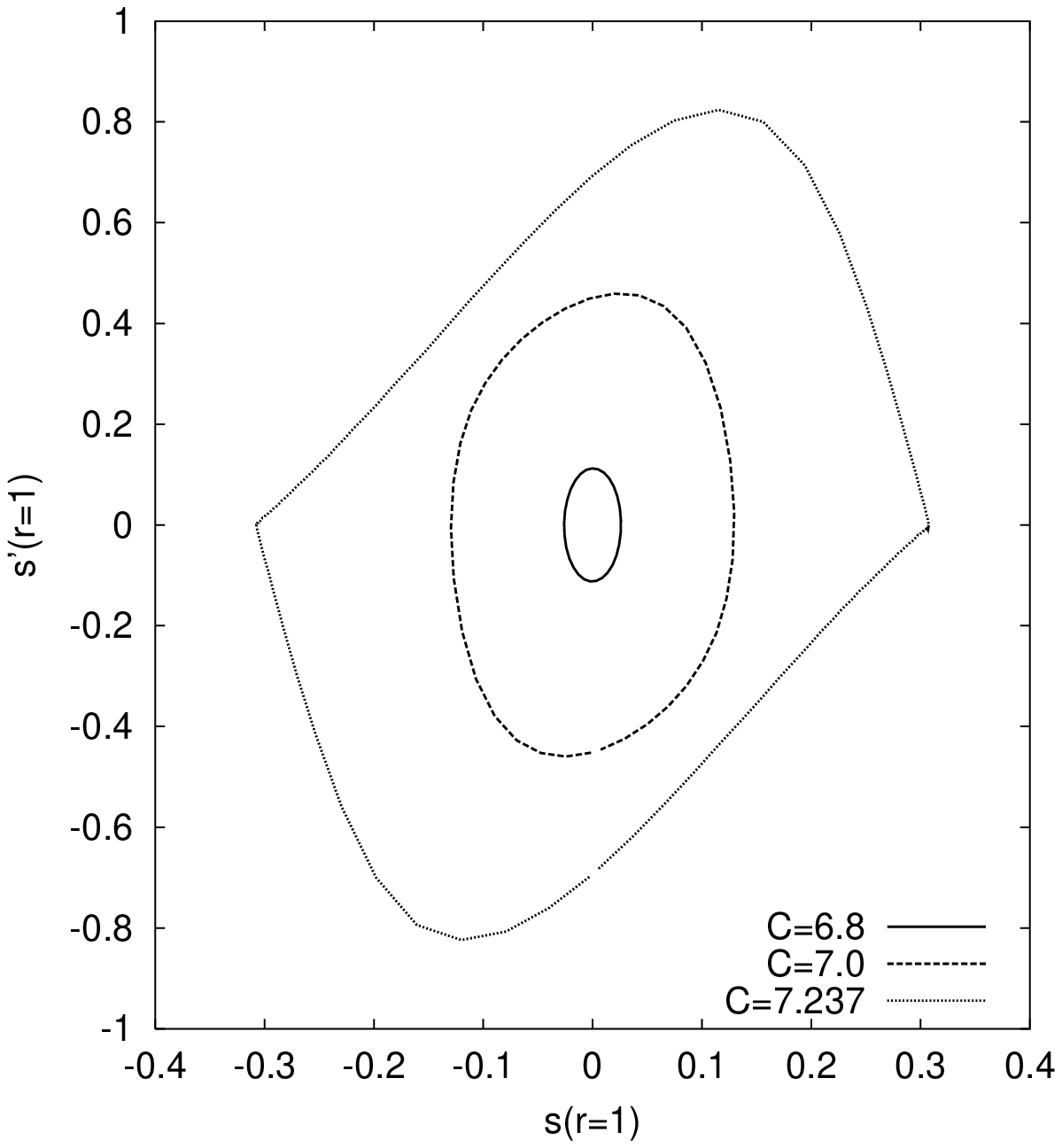}
\end{tabular}
\caption{Phase space trajectories for the van der Pol oscillator and
the dynamo with increasing $\mu$ and $C$, respectively.}
\end{figure}

\subsection{Self-tuning into reversal prone states}

Up to now we have only seen how a
dynamo which is oscillatory in the
kinematic and slightly supercritical regime changes,
via increasingly anharmonic
relaxation oscillation, into a steady dynamo
for higher degrees of supercriticality.
However, in the latter regime, we can learn
much more when we analyze
the spectrum of the dynamo operators
belonging to the
actual quenched $\alpha(r)$ profiles in the
saturated state. This will help us to
understand the disposition of the (apparently steady)
dynamo to undergo reversals under
the influence of noise.

For this purpose we have shown in Fig. 5 the
growth rates (left) and the frequencies (right) of the
dynamo
in the (nearly) kinematic regime (for C=6.78) and for
the quenched $\alpha$ profiles
in the saturated regime with 
increasing values of $C$.
Actually, the curves result from scaling the actual quenched 
$\alpha$ profiles
with an  artificial pre-factor $C^*$. This artificial scaling
helps to identify the position of the 
actual eigenvalue (corresponding to $C^*=1$) relative
to the exceptional point.
For $C=6.78$, we see that the first and second eigenvalue merge
at a value of $C^*$ at which the growth rate is less than zero.
At $C^*=1$ it is
evidently an oscillatory dynamo. However, already
for $C=8$ the exceptional
point has moved far above the zero growth rate line.
Interestingly, for even higher
values of $C$ the exceptional
point and the nearby local maximum return close 
to the zero growth rate line.
It can easily be
anticipated that  steady dynamos
which are characterized by a stable fixed point
become increasingly unstable with respect
to noise. More examples of this self-tuning mechanism
can be found in a preceding paper (Stefani et al 2006a).
\begin{figure}
\begin{center}
\epsfxsize=12cm
\epsffile{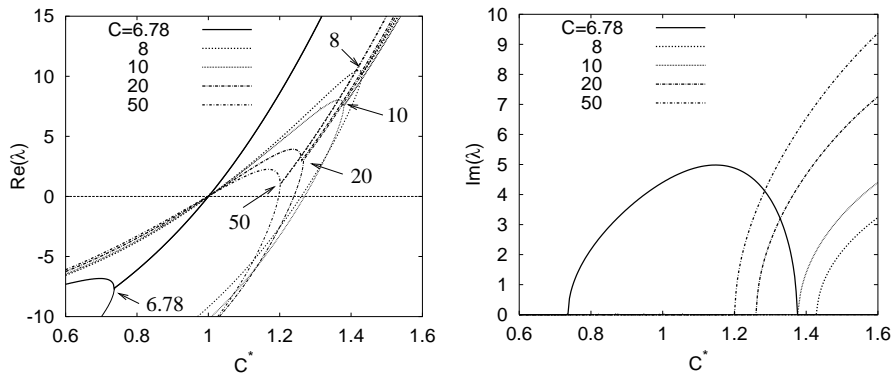}
\end{center}
\caption{Spectral properties of the nearly kinematic ($C=6.78$)
and of the saturated $\alpha(r)$ profiles (for $C=$8, 10, 20, 50) which
result from the chosen $\alpha_{kin}(r)=1.916 \cdot C  \cdot (1-6 \; r^2+5 \; r^4)$.
The scaling with the artificial factor $C^*$  helps to identify the
actual eigenvalue (at $C^*=1$) in its relative position to the
exceptional point. Note that for highly supercritical $C$ the
exceptional point moves close to the zero growth rate line.}
\end{figure}

\subsection{Reversals as noise induced relaxation oscillations}

In this section, we will study highly supercritical dynamos
under the influence
of noise. We will vary the parameters $C$ and $D$
and check their
influence on the time scale and the shape of reversals.
Then we will compare the phase space trajectories
during numerical reversals with
those of paleomagnetic ones. Although the
dynamo is not anymore in
the oscillatory regime
the typical features of relaxation oscillation
re-appear during the
reversal.
Figure 6 shows typical magnetic field series for a
time interval of 100 diffusion times which would
correspond to approximately 20 Myr
in time units of the real Earth.
Not very surprisingly, the increase of noise leads to an
increase of the reversal rate. The two documented
values of $C$
indicate also a positive correlation of dynamo
strength and
reversal frequency, although this
dependence is not always monotonic as was shown in
(Stefani et al. 2006a).
\begin{figure}
\begin{center}
\epsfxsize=12cm
\epsffile{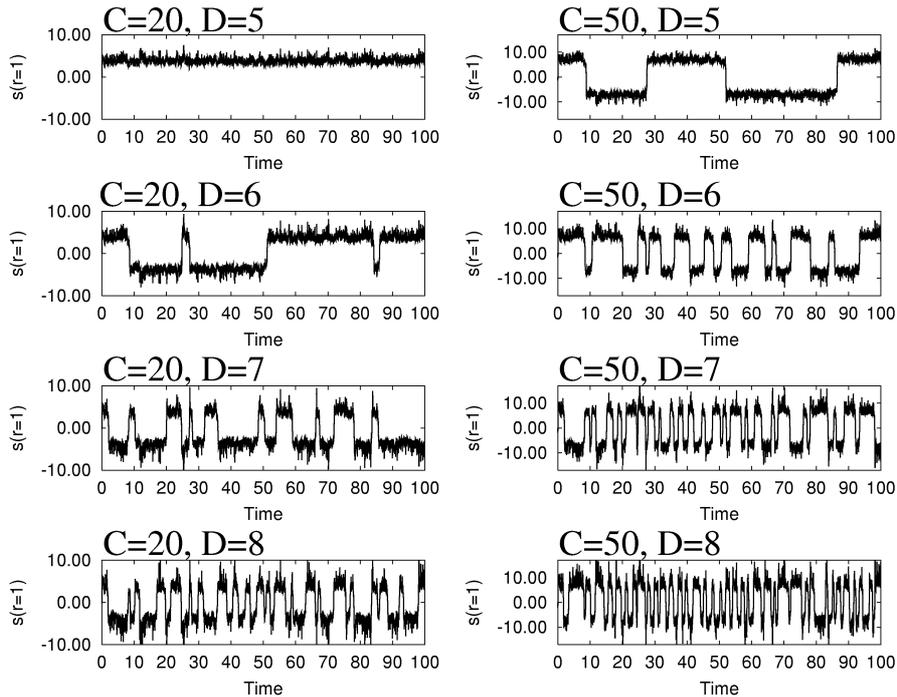}
\end{center}
\caption{Time series for various values of $C$ and $D$ for
the kinematic profile $\alpha_{kin}(r)=1.916  \cdot C  \cdot (1-6 \; r^2+5 \; r^4)$.}
\end{figure}

In Fig. 7 we compare now paleomagnetic
time series during reversals with time series resulting form 
the $\alpha^2$ model on one side and from the Hoyng-Duistermaat 
model on the other side.
Figure 7a represents the virtual axial dipole moments (VADM)
during the last five reversals as they were
published by Valet {\textit et al.} (2005).
Actually, the data points in Fig. 7a were extracted from Fig. 4
of (Valet {\textit et al.} 2005).
One can clearly see the strongly asymmetric
shape of the curves with a slow dipole decay that takes
around 50 kyr
followed by  fast dipole recovery taking approximately 5- 10 kyr.
Since the curves represent an average over many site samples
the VADM does not go exactly to zero at the reversal point.
It is useful to take an average over these five curves in
order to compare it with the time series following
from various  numerical models.
\begin{figure}
\begin{center}
\epsfxsize=14cm
\epsffile{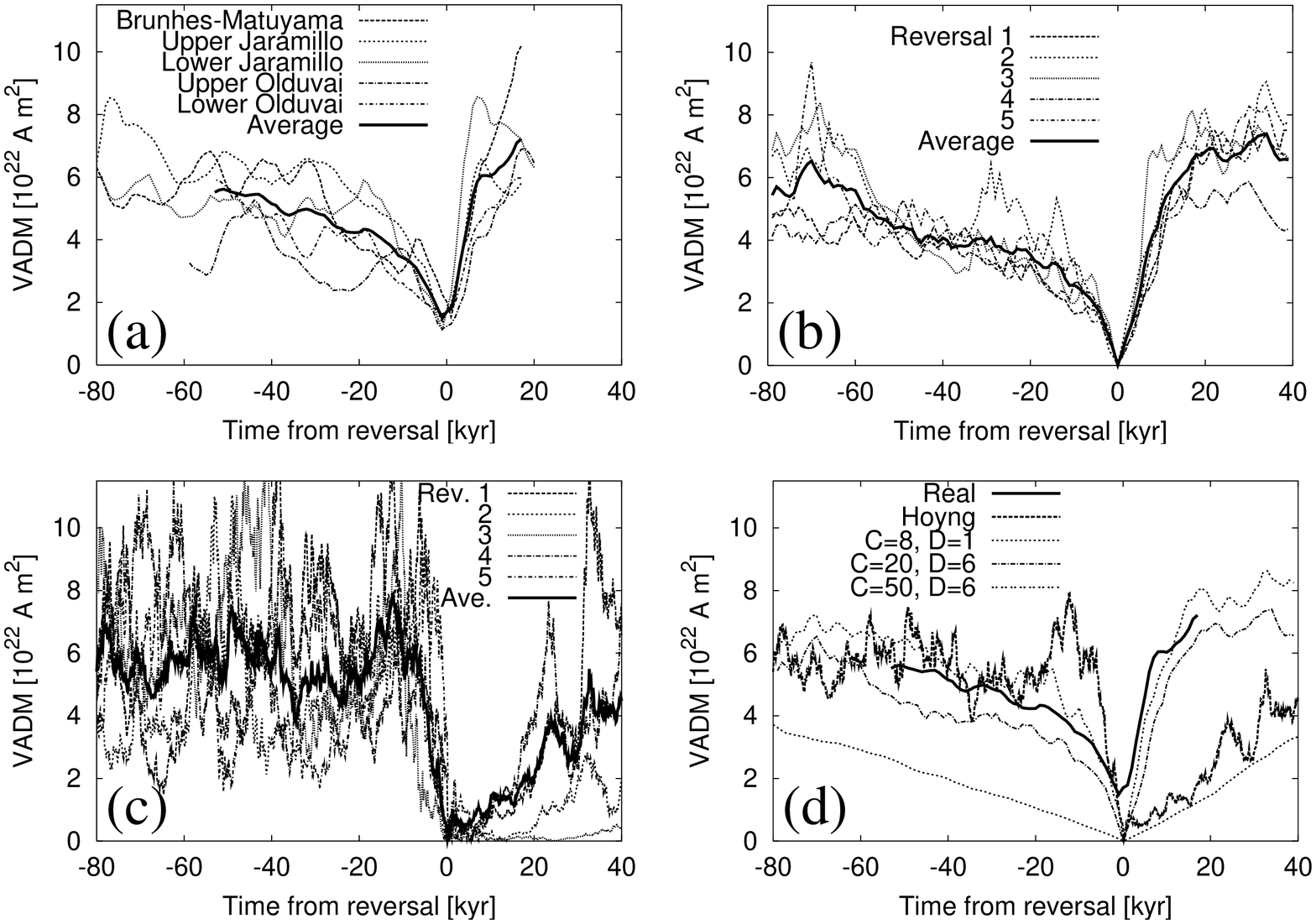}
\end{center}
\caption{Comparison of paleomagnetic reversal data and numerically
simulated ones. (a) Virtual axial dipole moment (VADM) during
the 80 kyr preceeding  and the 20 kyr following a polarity transition
for five reversals from the last 2 million years 
(data extracted from (Valet {\textit et al.} 2005)),
and their average. (b) Five typical reversals resulting from the dynamo model
with $\alpha_{kin}(r)=1.916 \cdot C  \cdot (1-6 \; r^2+5 \; r^4)$
for $C=20$, $D=6$, and their average.
(c) Five typical reversals resulting from the model of Hoyng and Duistermaat,
and their
average. Note that the time scale in this model has been chosen in such
a way that it becomes comparable to geodynamo reversals.
(d) Comparison of the average curves from (a), (c), and (b), complemented
by two further examples with $C=8$, $D=1$ and $C=50$, $D=6$.
The field scale for all the numerical
curves has been fixed in such a way that the intensity in the
non-reversing periods matches approximately the observed values.}
\end{figure}

The comparison of the real data with the numerical time
series for $C=20$, $D=6$
in Fig. 7b shows a nice correspondence.
Quite in contrast to this, the time
series of the Hoyng-Duistermaat model show a wrong asymmetry (Fig. 7c).
We compare the averages of real data, of our model results, and
of the Hoyng-Duistermaat model results in Fig. 7d.
Apart from the slightly supercritical
case $C=8, D=1$ which
exhibits a much to slow magnetic field evolution, the other
examples with $C=$20, 50 and  $D=6$ show very realistic time
series with the typical slow decay and  fast recreation.
As noted above, the fast recreation results from the fact 
that in a small interval during the transition the dynamo 
operates with a nearly
unquenched $\alpha(r)$ profile which yields, in case that
the dynamo is strongly supercritical,
very high growth rates.

With view on the relaxation oscillation property it may also
be instructive to show the phase space trajectories in Fig. 8.
In the curves of the real data, we have left out the data points
very close to the sign change since these are not very reliable.
Apart from this detail, we see in the paleomagnetic data and the
numerical data the typical asymmetric shape of
the phase space trajectories.
\begin{figure}
\begin{center}
\epsfxsize=14cm
\epsffile{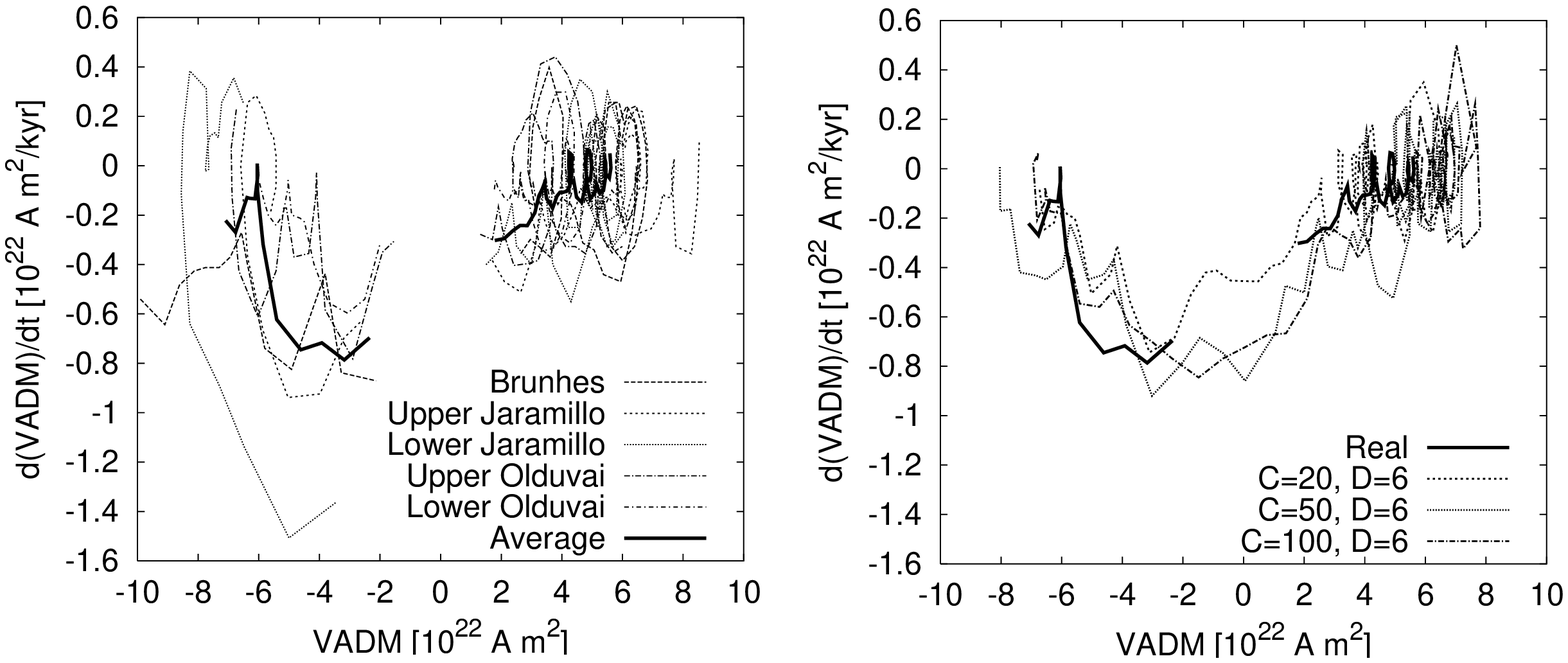}
\end{center}
\caption{Phase space trajectories of real and numerical reversals.
For the real data we have left out  three points close to the
very sign change position which are not very
reliable.}
\end{figure}

\section{Clustering properties of reversal sequences}
The question whether the time intervals between reversals
are governed by a Poisson process or not has been discussed by
several authors (McFadden 1984, Constable 2000).
The main difficulty in this approach is that, due to the poor statistics of the real data, 
the distribution of the time intervals between successive events is not clearly
defined. In particular, it is not obvious to distinguish 
between an exponential distribution (that would be the case for a random, Poisson process)
and a power-law process (indicating presence of correlations). 
Moreover, Constable (2000) has shown that the mean reversal rate is changing
in the course of time, which would correspond to a
non-stationary Poisson process with a time dependent rate function.
This phenomenology could generate a power-law statistics for the inter-event times, 
even in absence of correlations. 
Only recently Carbone {\textit et al.} (2006) were able to
prove the non-Poissonian character of the reversal process by applying a criterion
to the reversal sequence which works also for non-local Poisson processes.
\begin{figure}
\begin{center}
\epsfxsize=14cm
\epsffile{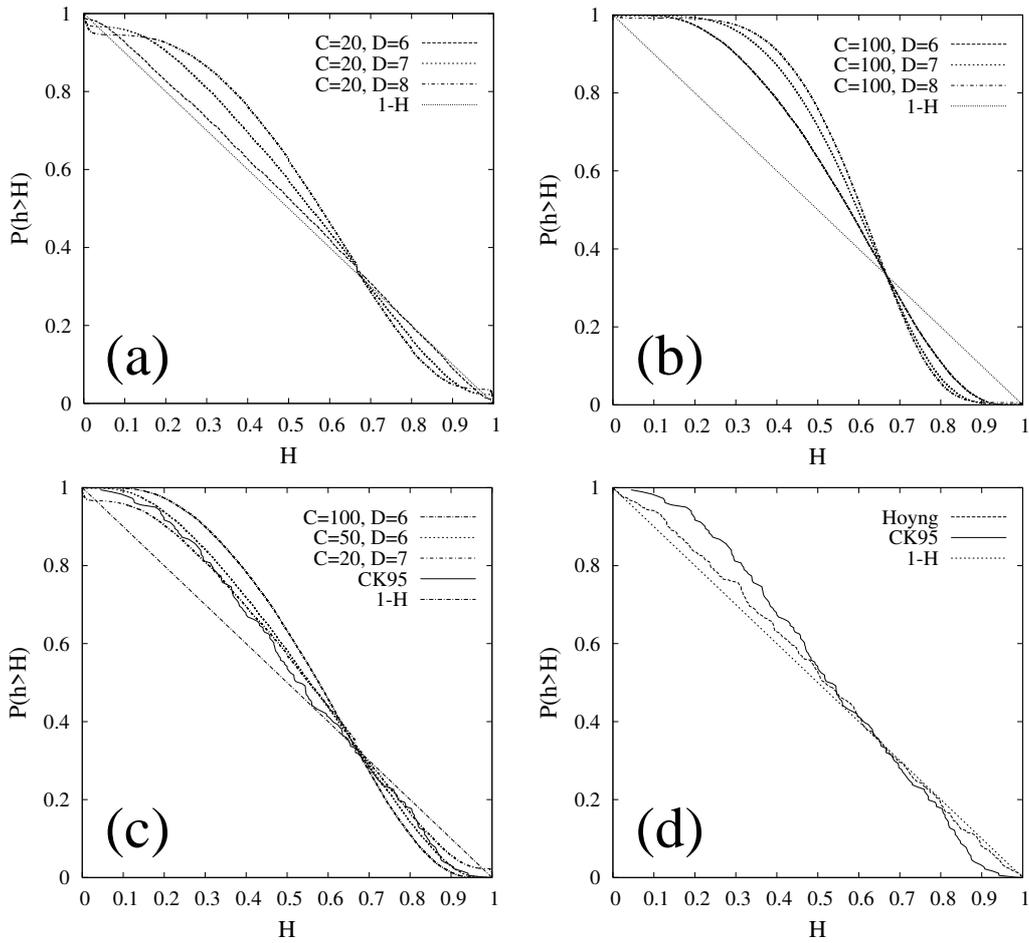}
\end{center}
\caption{The dependence $P(h>H)$ on $H$ for various numerical
models and for the paleomagnetic date take from Cande and Kent (1995).
For the parameter choice $C=20$, $D=6$ and $C=50$, $D=6$ we see
a nearly perfect agreement with the real data. In contrast,
the reversal data of
the Hoyng model are much closer to a Poisson process.}
\end{figure}

A central quantity in their analysis is
\begin{eqnarray}
H(\delta t, \Delta t)&=&2 \delta t/(2 \delta t+\Delta t)
\end{eqnarray}
wherein, for a given reversal instant
$t_i$, $\delta t$ is defined as
the minimum of the
preceding and subsequent time interval:
\begin{eqnarray}
\delta t&=&\min\{t_{i+1}-t_i; t_i-t_{i-1}\} \; .
\end{eqnarray}
$\Delta t$ is then the ``pre-preceding" or
"sub-subsequent"
time interval, respectively, according to
\begin{eqnarray}
\Delta t&=&t_{i+2}-t_{i+1} \;\; \mbox{or} \;\; t_{i-1}-t_{i-2} \; .
\end{eqnarray}
The meaning of $H(\delta t, \Delta t)$ is quite simple:
if reversals are clustered
then the $\Delta t$ (which follows or precedes the
$\delta t$ which is, by definition (10), assumed to be small)
will also be small, and since $\Delta t$
appears in the denominator of Eq. (9),
$H(\delta t,\Delta t)$ will be comparably large.
In contrast, if the reversal sequence is governed by voids,
then $\Delta t$ will be rather large and
$H(\delta t,\Delta t)$ will be rather small.
It can easily be shown that, for a sequence of random events, described by a
Poisson process, the values of $H(\delta t, \Delta t)$ must be
uniformly distributed in $[0,1]$.
The interesting thing with this formulation is that,
because of its local character, it
indicates clustering or the appearance of voids
even in the case of inhomogeneous Poisson processes
(with time dependent rate function).

The sequence of paleomagnetic reversal data was shown
to have a significant tendency to cluster (Carbone {\textit et al.} 2006).
In a follow-up paper (Sorriso-Valvo {\textit et al.} 2006),
various simplified dynamo models were examined with respect
to their capability to
describe this clustering process. In particular it turned out
that the $\alpha^2$ dynamo model showed the right clustering property,
although in a more pronounced way than
the paleomagnetic data (Cande and Kent 1995).

In this section we analyze this behaviour
a bit more in detail by checking the
quantity H for a variety of values $C$ and $D$.
In order to better emphasize and better visualize the 
shape of the $H$ distribution function, we show here the
corresponding surviving function, namely the probability $P(h>H)$. 
In the Poisson (or local Poisson) case, that is uniform distribution of $H$, 
the surviving function would depend linearly on $h$, namely $P(h>H)=1-h$.
The results are shown in Fig. 9. The dependence of $P(h>H)$ on H is shown
for different values of $D$ from $C=20$ (Fig. 9a) and
$C=100$ (Fig. 9b). In Fig. 9c, we compare the best curves
for $C=20$, 50, and 100 with the paleomagnetic data
(Cande and Kent 1995). Especially for $C=20$, $D=7$ and
$C=50$, $D=6$ we observe a nearly perfect agreement,
indicating that the temporal distribution of the reversals captures the
real data features, including their tendency to cluster.
In contrast to this, the model of Hoyng and Duistermaat shows more or
less a Poisson behaviour (Fig. 9d).
Note that the discrepancy of the results presented here for the Hoyng model from those 
reported in Carbone et al. (2006) and in Sorriso-Valvo et al. (2006), 
are due to the removal of the excursion-like events.

\section{The role of the inner core}
In this section we will study the influence of an
inner core within the framework of our simple
model. The usual picture of the role of the inner core
was expressed in two papers by Hollerbach and
Jones (1993, 1995). Basically, one expects that
the conducting inner core impedes the occurrence of
reversals
by the effect that the magnetic field evolution is
governed
there by the
long diffusion time scale and
cannot follow the magnetic field evolution in the
outer core which is governed by much shorter
time scales of convection. Finally this will lead to
a dominance of excursion over reversals
(Gubbins 1999).

In the following we will check if and
how this
simple picture translates into our model.
To begin with, we analyze the magnetic field evolution
for a family of
kinematic $\alpha$ profiles
\begin{eqnarray}
\alpha_{kin}(x)=\frac{1.914 \cdot C}{1+\exp{[(x_0-x)/d]}}\left[1.15-6 \left(\frac{x-x_0}{1-x_0}\right)^2+
5\left(\frac{x-x_0}{1-x_0}\right)^4 \right] \; .
\end{eqnarray}
Basically this is a similar
profile as considered before
but the denominator $1+\exp{\left[(x_0-x)/d\right]}$ makes it vanishing
in the inner core region with radius
$x_0$ (including a smooth transition region
of thickness $d$
which is simply used for numerical reasons).
\begin{figure}
\begin{center}
\epsfxsize=14cm
\epsffile{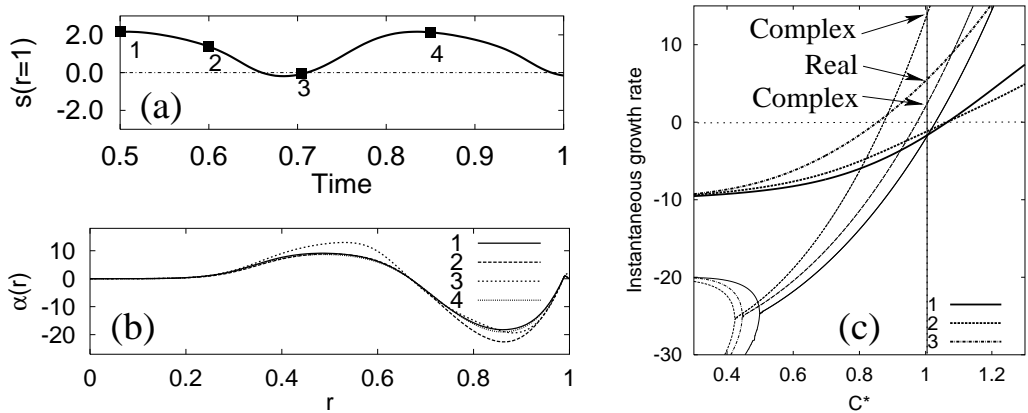}
\end{center}
\caption{The spectral explanation of excursions: Time evolution (a),
instantaneous $\alpha(r)$ profiles (b), and
instantaneous growth rates  (c) for the kinematic
$\alpha(x)$ according to Eq. (12)
with special values $C=20$, $x_0=0.35$ and $d=0.05$.}
\end{figure}
For this model, with the special choice $C=20$, $x_0=0.35$ and
$d=0.05$, Fig. 10a shows the time evolution of $s(x=1)$.
In contrast to all  models considered before, we get
now an oscillation around
a non-zero mean value (which was already observed by
Hollerbach and Jones
(1995)). Analogously as reversals
can be traced to the transition between
steady and oscillatory modes, the new ``excursive''
behaviour can also be explained in terms of the spectral properties of
the underlying dynamo operator. To see this we consider in Fig. 10b the
instantaneous $\alpha$ profiles at the instants 1, 2, 3, 4 indicated in
Fig. 10a.
For these instantaneous $\alpha$ profiles we show the first
three eigenvalues in Fig. 10c, again in dependence on the artifical
scaling parameter $C^*$. In this picture an excursion occurs as follows:
First, at instant 1, the dominant eigenvalue is the complex one
resulting from the
merging of the second and third radial eigenmode. Its real part
is clearly larger than the real part of the first eigenvalue.
Being thus governed by a complex eigenvalue, the system starts to
undergo a reversal. However,
during this attempted oscillation the non-linear
back-reaction becomes weaker (instant 2). Hence
$\alpha$ becomes less quenched and we end
up with a new instantaneous dynamo  operator whose first
real eigenvalue becomes dominant and whose second
oscillatory
branch becomes subdominant.
Therefore, the system is governed now by a real
and positive eigenvalue and the magnetic field
increases again before it was able to complete the sign change.
Later, the system returns to the state with
stronger $\alpha$ quenching
were the oscillatory mode becomes dominant again  and so on.

That way, we have traced back excursions to the
intricate ordering of the real first eigenvalue and
the complex merger of the second and the
third eigenvalue. The deep reason for this complicated behaviour
has
been indicated in the paper
by G\"unther and Kirillov (2006), although for the
case of zero boundary conditions which
changes the spectrum drastically. In terms of the
total $\alpha(r)$ profile, an inner core of a
certain radius will contribute in form
of higher radial harmonics.
Splitting the induction equation into a coupled
equation system for different radial harmonics
(here Bessel functions) we observe a resonance phenomenon.
Radial $\alpha$ profiles with higher harmonics in radial direction
will also favour
radial magnetic field modes with comparative wavelengths.
Roughly speaking, an $\alpha(r)$ profile with one sign change
along the radius (i.e. of the
form $(1-6x^2+2 -5x^4)$) will favour
the second eigenmode whose larger growth rate will tend
to merge with that of the first eigenmode. Including a core,
however, the third harmonic will be favoured and will
merge with the second
eigenfunction and so on.

Now let us switch to a core model with higher supercriticality.
In Fig. 11 we considered the modified (and,
admittedly, somewhat tuned)
time dependent kinematic $\alpha(r)$ profile:
\begin{eqnarray}
\alpha_{kin}(x)&=&\frac{1.914 \cdot C}{1+\exp{\left[(x_0-x)/d\right]}}\times  \nonumber \\
&&\times \left[0.7+3 \cdot x_0/1.914-6 \left(\frac{x-x_0}{1-x_0}\right)^2+4.5
\left(\frac{x-x_0}{1-x_0}\right)^4 \right]
\end{eqnarray}
wherein the noise respects the core geometry in an appropriate  manner
\begin{eqnarray}
\Xi(r,\tau)=\frac{1}{1+\exp{\left[(x_0-x)/d\right]}}\left(\xi_1(\tau) +\xi_2(\tau) \; r^2 +\xi_3(\tau) 
\; r^3+\xi_4(\tau) \; r^4 \right) \; \; .
\end{eqnarray}
We carry out a simulation of this model, with $C=50$, $D=6$, and $d=0.05$,
by increasing the inner core radius $x_0$ every 10 diffusion times
by 0.05 (see Fig. 11b). The result can be seen in Fig. 11a.
\begin{figure}
\begin{center}
\epsfxsize=14cm
\epsffile{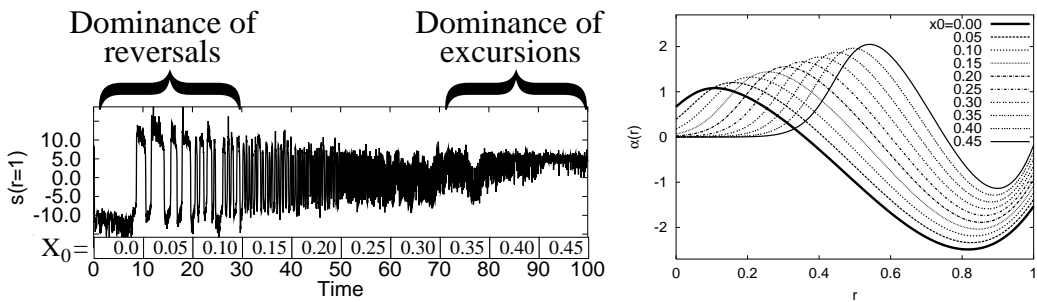}
\end{center}
\caption{The influence of a growing core on the reversal rate and
the arising predominance of excursions.}
\end{figure}

The first observation is that the role of the inner core
is quite complicated. Starting from $x_0=0$ we see that
with increasing $x_0$ the
reversal frequency increases drastically.
This is in clear contradiction to
the Hollerbach and Jones picture.
The second observation is that for larger $x_0$
the inner core indeed has a tendency to favour
excursions in comparison with reversals which
confirms again the Hollerbach and Jones view.
At least we can see that the role of the inner
core is much more complicated since the
selection of modes which are coupling at an exceptional
point depends very sensitively on its radius.

\section{Towards more realistic models}

We will conclude our study with an ``excursion''  into
more realistic dynamo models. As we have seen, the
time characteristics and the asymmetry
of the real reversal process is well recovered
within a simple dynamo model that
requires only a dynamo operator with a  local maximum of the growth rate
and a nearby exceptional point where the two real eigenvalues
merge and continue as oscillatory mode. This is a very unspecific
requirement which does not rely on specific assumptions on the
flow structure in the core. Nevertheless, it would be
nice to see if and how this picture
translates into more realistic dynamo models.

In the model of Giesecke (2005a), which includes the
North-South asymmetry
of $\alpha$ by virtue  of a $\cos(\theta)$ term, quite similar
reversal sequences were
observed. Actually, in contrast to our model,
Gieseckes model assumes
a turbulent resistivity which is approximately 10
times larger than the molecular one.
We believe that this necessity would disappear when
going over from
slightly to highly
supercritical dynamos. In this respect it is
instructive to compare again the two
curves for $C=8$ and $C=20$ in Fig. 7d which are
characterized by very
different decay and growth times.

Apart from this detail, it is worthwhile to look a bit deeper into the spectral
property of Gieseckes dynamo  model. For that purpose we have analyzed
the spectrum of the dynamo operator for functions
of $\alpha(r,\theta)$ according to
\begin{eqnarray}
\alpha_{kin}(r,\theta)=Rm \; \cos( \theta) \; \sin \left[2 \pi (x-x_0)/(1-x_0)\right] \; .
\end{eqnarray}
Figure 12 shows the eigenvalues of the
first five axisymmetric modes (m=0) and the first three non-axisymmetric
modes (m=1) for different values of the inner core size $x_0$. The
growth rates and frequencies  result from
an integral equation solver which was 
documented in previous papers
(Stefani {\textit et al.} 2000, Xu {\textit et al. 2004a},
Xu {\textit et al.} 2004b, Stefani {\textit et al.} 2006c).
The most important point which is changed by the inclusion
of the $\cos(\theta)$ term is that now  different
$l$-modes are not independent
anymore. Hence it may happen that modes with neighbouring  $l$ merge at
an exceptional point which was forbidden in the spherically symmetric model
considered so far. Therefore, the
scheme of mode coupling becomes even more complicated as before and one
might expect
a higher sensitivity with respect to changes of
the inner core radius $x_0$.

Restricting our attention to the $m=0$ modes, we observe for
$x_0=0$ an  immediate coupling of the modes 2 and 3, which
split off again close
to $Rm=4$. The next coupling occurs close to $Rm=11.5$, but the resulting
low lying mode seems not
relevant for the reversal process. For $x_0=0.2$, the modes 2 and 3
are coupled from the
very beginning, but the mode 1 and 4 couple also close to $Rm=13$.
For $x_0=0.35$, which is approximately the case for the Earth,
we see an immediate coupling of the modes 2 and 3 on one side and of
the modes 4 and 5 on the other side. The latter split off again
close to $Rm=16$, but shortly after this the mode 4 couples again with
the mode 1. For $x_0=0.5$, the low modes show no coupling at all, only
the mode 5 couples to the mode 6 which is, however, not shown in the
plot.
\begin{figure}
\begin{center}
\epsfxsize=12cm
\epsffile{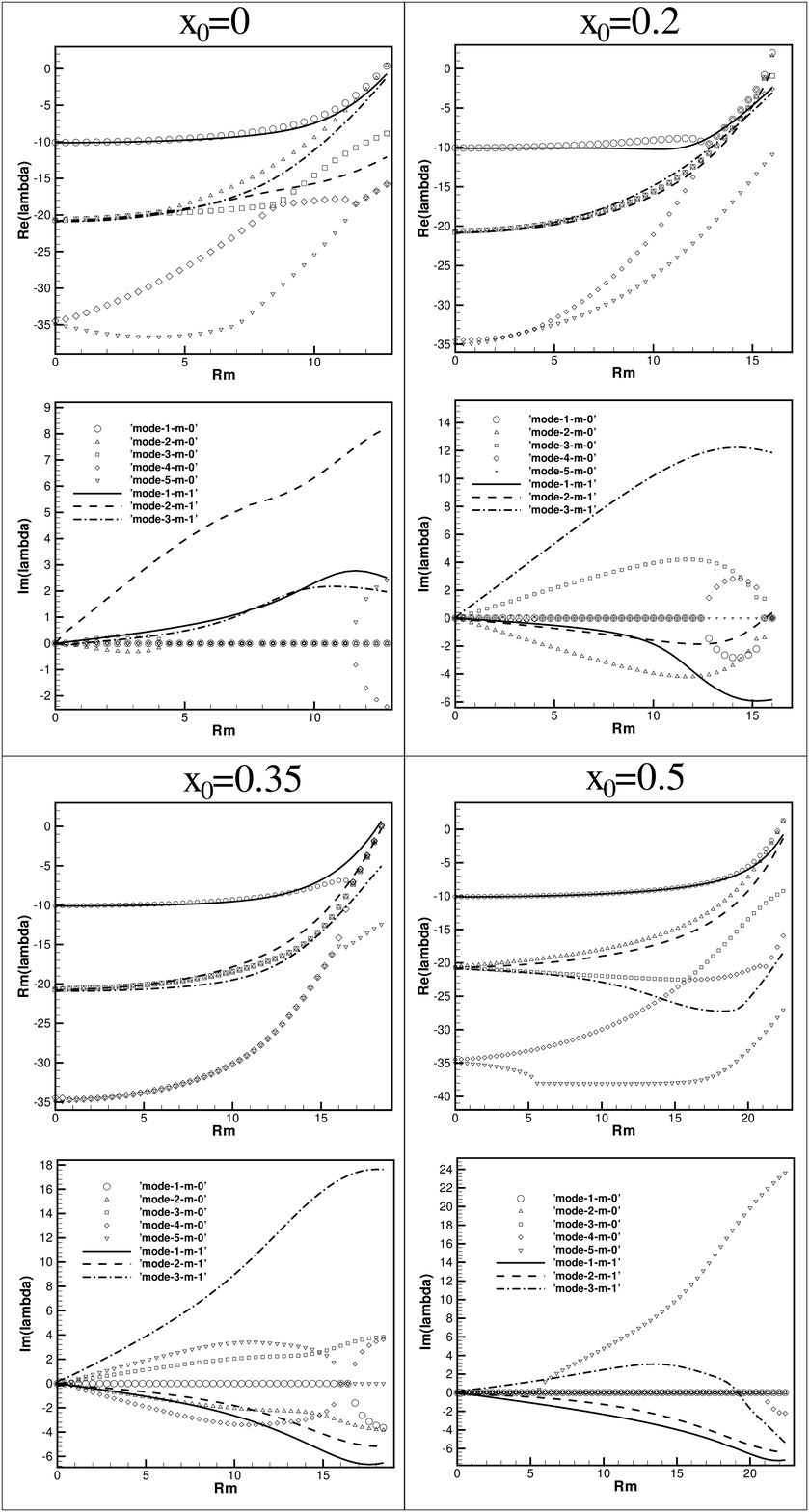}
\end{center}
\caption{Spectra for  $\alpha^2$ dynamos respecting the North-South
asymmetry by  virtue of a $\cos(\theta)$
term, in dependence on the inner core size $x_0$.}
\end{figure}
In summary, this gives a complicated scheme of mode coupling and
exceptional points which is quite sensitive to the value of $x_0$.
What is missing at the moment is a detailed study of the
spectral properties
of the
dynamo operator in the {\it saturated regime} which must be left to
future work. We had indicated
in Fig. 5 (and substantiated in more detail in (Stefani {\textit et al.}
2006b)) that highly supercritical
dynamos have a tendency to self-tune into a state in which
they are prone to reversals.
Hence Fig 12 can be quite misleading  when it comes
to the exact position of the exceptional points of the
dominant mode. And, of course, one should not forget
that the nearly degeneration between $m=0$ and $m=1$ will also
be lifted
if a tensor structure of $\alpha$ and further mean-field
coefficients come into play as it is typical in more
realistic mean-field models of the geodynamo
(Schrinner {\textit et al.} 2005, 2006).

\section{Summary and speculations}

We have checked the capability of  a simple $\alpha^2$ dynamo model
with spherical symmetric $\alpha$
to explain the typical time-scales and the
asymmetric shape of Earth's magnetic field reversals.
The results are
quite encouraging, in particular when comparing the paleomagnetic  and
the numerical reversal data
in Figs. 7 and 8. In our view, the interpretation of reversals as
noise induced relaxation oscillations seems quite natural.
This mechanism of reversals  relies on the existence of
a local growth rate maximum and a nearby exceptional point of the
spectrum of the dynamo operator in the saturated regime.

The clustering properties which were recently shown to be
present in real paleomagnetic data (Carbone \textit{et al.} 2006)
were also observed in the model. Both the typical time scales
of reversals and the degree of deviation from Poissonian statistics
indicate that the Earth dynamo works in a regime of high, although not
extremely high, supercriticality.

We conclude this paper with a speculation on a possible connection of
the inner core radius with the long term behaviour of the  reversal rate,
including the occurrence of superchrons.
It starts from the observation that
the reversals rate of the last 160 Myr  (see Merrill  \textit{et al.} 1996) 
suggests
a time scale of long term variation in the order of  100 Myr.
If one would take into account the Kiaman superchron, and possibly the
(up to present unsettled) third superchron discovered by Pavlov and Gallet, one
might hypothesize a process with a period of approximately
200 Myr. What could be the nature of such a  periodicity (if there were
any)? The first idea that comes to mind is the connection with slow
mantle convection processes that influence, by virtue of
changed  core-mantle boundary conditions,
the heat transport in the core, hence the dynamo strength and ultimately the
reversal rate.

But what about the inner core? Is there a possibility that such a
200 Myr variability (or periodicity?) be
related to a resonance phenomenon of the spectral properties of the dynamo
operator for increasing inner core size? What we have seen at least
 is a strong influence of the inner core size on the
selection of the two eigenmodes which merge
together at an exceptional point into an oscillatory mode. This selection, in turn,
has a strong influence on the reversal rate. A most important criterion  for testing
such a hypothesis  is the
age of the inner core. Unfortunately, the data on this point vary wildly in the
literature, ranging  from values of 500 Myr until 2500 Myr with a most likely value
around  1000 Myr (Labrosse 2001).
It is not excluded that a typical periodicity of some 200 Million years could result from
an inner core growing on these time scales. At least it seems
worth to test this hypothesis by further analysing
the spectra of realistic and highly supercritical  dynamo models in the saturated regime.

\section*{Acknowledgments}
This work was supported by Deutsche Forschungsgemeinschaft
in frame of SFB 609 and under Grant No. GE 682/12-2.


\begin{thebibliography}{11}


\bibitem{Aubert}Aubert, J. and Wicht, J., Axial versus equatorial dipolar dynamo models
with implications for planetary magnetic fields. \textit{Earth. Plan.
Sci. Lett.}, 2004, \textbf{221}, 409-419.
\bibitem{CANDEKENT} Cande, S. C. and Kent, D. V., Revised calibration
of the geomagnetic polarity timescale for the late Cretacious and Cenozoic.
\textit{J. Geophys. Res. - Solid Earth}, 1995, \textbf{100}, 6093-6095.
\bibitem{CARBONE} Carbone, V., Sorriso-Valvo, L., Vecchio, A., Lepreti, F.,
Veltri, P., Harabaglia, P. and  Guerra I,
Clustering of polarity reversals of the geomagnetic field. \textit{Phys. Rev. Lett.}, 2006,
\textbf{96}, Art. No. 128501.
\bibitem{CHIGI}Childress, S. and Gilbert, A. D.,
{\it Stretch, Twist, Fold: The Fast Dynamo}. 1995,
(Springer, Berlin).
\bibitem{CLEMENT}Clement, B. M., Dependence of
the duration of geomagnetic polarity reversals on site latitude.
\textit{Nature}, 2004, \textbf{428}, 637-640.
\bibitem{CONSTABLE}Constable, C., On rates of occurrence of geomagnetic reversals.
\textit{Phys. Earth Planet. Inter.}, 2000, \textbf{118},  181-193.
\bibitem{COX} Cox, A., Length of geomagnetic polarity intervals. \textit{J. Geophys. Res.},
1968, \textbf{73}, 3247-3259.
\bibitem{GLRO}Glatzmaier, G. A. and  Roberts,  P. H.,
A 3-dimensional self-consistent computer simulation of a geomagnetic field
reversal. \textit{Nature},
1995, \textbf{377},  203-209.
\bibitem{Giesecke-a}Giesecke, A., R\"{u}diger, G. and Elstner, D., Oscillating $\alpha^{2}$-dynamos
and the reversal phenomenon of the global geodynamo. \textit{Astron.
Nachr.}, 2005a, \textbf{326}, 693-700.
\bibitem{Giesecke-b}Giesecke, A., Ziegler, U. and R\"{u}diger, G., Geodynamo $\alpha$-effect
derived from box simulations of rotating magnetoconvection. \textit{Phys.
Earth Planet. Inter.}, 2005b, \textbf{152}, 90-102.
\bibitem{Grote}Grote, E. and Busse, F.H., Hemispherical dynamos generated by convection
in rotating spherical shells. \textit{Phys. Rev. E}, 2000, \textbf{62},
4457-4460.
\bibitem{Gubbins}Gubbins, D., Barber, C.N., Gibbons, S. and Love, J.J., Kinematic dynamo
action in a sphere. II Symmetry selection. \textit{Proc. R. Soc. Lond.
A}, 2000, \textbf{456}, 1669-1683.
\bibitem{GubbinsLove}Gubbins, D. and Love, J. J., Preferred VGP paths during
geomagnetic
polarity reversals: Symmetry considerations.
\textit{Geophys. Res. Lett.}, 1998, \textbf{25}, 1079-1082.
\bibitem{Gubbinsexcursions}Gubbins, D., The distinction
between geomagnetic excursions and reversals.
\textit{Geophys. J. Int.}, 1999, \textbf{137}, F1-F3.
\bibitem{GubbinsGibbons}Gubbins, D. and Gibbons, S.,
Three-dimensional dynamo waves in a sphere.
\textit{Geophys. Astrophys. Fluid Dyn.}, 2002, \textbf{96}, 481-498.
\bibitem{GUKI} G\"unther, U. and Kirillov, O., A Krein space related perturbation
theory for MHD $\alpha^2$ dynamos and resonant unfolding of diabolic
points. \textit{J. Phys. A}, 2006, \textbf{39}, 10057-10076.
\bibitem{HELLER} Heller, R., Merrill, R. T. and McFadden, P. L.,
The two states of paleomagnetic field intensities for the past
320 million years. \textit{Phys. Earth Planet. Inter.}, 2003, \textbf{135},
211-223
\bibitem{HOLLERBACH93} Hollerbach, R. and Jones, C. A., Influence of the
inner core on geomagnetic fluctuations and reversals.
\textit{Nature}, 1993, \textbf{365}, 541-543.
\bibitem{HOLLERBACH95} Hollerbach, R. and Jones, C.A., On the
magnetically stabilizing role of the Earth's inner core.
\textit{Phys. Earth Planet. Inter.}, 1995, \textbf{87}, 171-181.
\bibitem{HOYNG}Hoyng, P, Ossendrijver, M. A. J. H.  and Schmidt, D.,
The geodynamo as a bistable oscillator. \textit{Geophys. Astrophys. Fluid Dyn.}, 2001,
\textbf{94}, 263-314.
\bibitem{HOYNGDUISTERMAAT}Hoyng, P.  and Duistermaat, J. J.,
Geomagnetic reversals and the stochastic exit problem.
\textit{Europhys. Lett.}, 2004, \textbf{68}, 177-183.
\bibitem{KRAUSE}Krause, F. and R\"adler, K.-H.,
{\it Mean-field Magnetohydrodynamics and Dynamo Theory}. 1980,
(Akademie-Verlag, Berlin).
\bibitem{LABROSSE} Labrosse, S., Poirier, J.-P. and LeMouel, J. L., The age of the
inner core. \textit{Earth Planet. Sci. Lett.}, 2001, \textbf{190}, 111-123.
\bibitem{MCFADDEN84}McFadden, P. L., Statistical tools for the
analysis of geomagnetic reversal sequences. \textit{J. Geophys. Res.}, 1984,
\textbf{89}, 3363-3372
\bibitem{MEINELBRANDENBURG} Meinel, R. and Brandenburg, A.,
Behaviour of highly supercritical $\alpha$-effect dynamos. \textit{Astron Astrophys.},
1990, \textbf{238}, 369-376.
\bibitem{Melbourne}Melbourne, I., Proctor, M. R. E. and Rucklidge, A. M., A heteroclinic
model of geodynamo reversals and excursions, in: \textit{Dynamo and
Dynamics, a Mathematical Challenge} (eds. P. Chossat, D. Armbruster
and I. Oprea), Kluwer, Dordrecht, 2001, pp. 363-370.
\bibitem{MERRILL} Merrill, R. T., McElhinny, M. W. and McFadden, P. L.,
{\it The magnetic field of the Earth}, 1996, (Academic Press: San Diego)
\bibitem{MININNI}Mininni, P. D., Gomez, D. O. and Mindlin, G. B.,
Stochastic relaxation oscillator model for the solar cycle.
\textit{Phys. Rev. Lett.}, 2000,  \textbf{85}, 5476-5479.
\bibitem{NARTEAU}Narteau, C., Private communication, 2005
 \bibitem{Olson}Olson, P., Christensen, U. and Glatzmaier, G.A., Numerical modelling
of the geodynamo: Mechanisms of field generation and equilibration.
\textit{J. Geophys. Res.}, 1999, \textbf{104}, 10383-10404.
\bibitem{PAVLOVGALLET}Pavlov, V. and Gallet, Y.,
A third superchron during early Paleozoic. \textit{Episodes}, 2005, \textbf{28}, No 2, 78-84.
\bibitem{PERRIN} Perrin, M. and Shcherbakov, V. P., Paleointensity of the Earth's
magnetic field for the paste 400 Ma: evidence for a dipole
structure during the mesozoic low. \textit{J. Geomagn. Geoelectr}, 1997,
\textbf{49}, 601-614.
\bibitem{Phillips}Phillips, C. G., Ph.D. Thesis, 1993, University of Sidney.
\bibitem{PONTIERI}  Pontieri, A., Lepreti, F., Sorriso-Valvo, L., Vecchio, A., Carbone, V.,
A simple model for the solar cycle. \textit{Solar Physics}, 2003,   \textbf{213}, 195-201.
\bibitem{Sarson}Sarson, G.R . and Jones, C.A., A convection driven geodynamo reversal
model. \textit{Phys. Earth Planet. Inter.}, 1999, \textbf{111}, 3-20.
\bibitem{SCHMIDT}Schmidt, D., Ossendrijver, M. A. J. H. and Hoyng P.,
Magnetic field reversals and secular variation in a bistable geodynamo
model. \textit{Phys. Earth Planet. Inter}, 2001, \textbf{125}, 119-124.
\bibitem{SCHRINNER1}Schrinner, M, R\"adler, K.-H., Schmitt, D., Rheinhardt, M.,
Christensen, U. R., Mean-field view on rotating magnetoconvection
and a geodynamo model. \textit{Astron. Nachr.}, 2005, \textbf{326},  245-249.
\bibitem{SCHRINNER2}Schrinner, M, R\"adler, K.-H., Schmitt, D., Rheinhardt, M.,
Christensen, U. R., Mean-field concept and direct numerical simulations of 
rotating magnetoconvection and the geodynamo. \textit{Geophys. Astrophys. Fluid
Dyn.}, 2006, submitted; Preprint: astro-ph/0609752.
\bibitem{SHCHERBAKOV}Shcherbakov, V. P., Solodovnikov, G. M. and Sycheva, N. K.,
Variations in the geomagnetic dipole during the past 400 million years
(volcanic rocks), \textit{Izvestiya, Phys. Solid Earth}, 2002, \textbf{38}, 113-119.
\bibitem{SORRISO-VALVO}Sorriso-Valvo, L., Stefani, F., Carbone, V.,
Nigro, G., Lepreti, F., Vecchio, A. and Veltri, P., A statistical
analysis of polarity reversals of the geomagnetic field.
\textit{Phys. Earth Planet. Inter}, 2006, submitted.
\bibitem{Stefani-e}Stefani, F., Gerbeth, G. and R\"{a}dler, K.-H., Steady dynamos in
finite domains: an integral equation approach. \textit{Astron. Nachr.},
2000, \textbf{321}, 65-73.
\bibitem{STEFANIGERBETH2003}Stefani, F. and Gerbeth, G.,
Oscillatory mean-field dynamos with a spherically symmetric, isotropic 
helical turbulence parameter $\alpha$. \textit{Phys. Rev. E}, 2003, \textbf{67}, Art. No. 027302.
\bibitem{Stefani2005}Stefani, F. and Gerbeth, G., Asymmetry polarity reversals, bimodal
field distribution, and coherence resonance in a spherically symmetric
mean-field dynamo model. \textit{Phys. Rev. Lett.}, 2005, \textbf{94},
Art. No. 184506.
\bibitem{Stefani2006a}Stefani, F., Gerbeth, G., G\"{u}nther, U. and Xu, M., Why dynamos
are prone to reversals. \textit{Earth Planet. Sci. Lett.}, 2006a,
\textbf{143}, 828-840.
\bibitem{Stefani2006b}Stefani, F., Gerbeth, G. and G\"{u}nther, U., A paradigmatic model
of Earth's magnetic field reversals. \textit{Magnetohydrodynamics},
2006b, \textbf{42}, 123-130.
\bibitem{Stefani2006c}Stefani, F., Xu, M., Gerbeth, G., Ravelet, F., Chiffaudel, A., Daviaud, F. and
Leorat, J., Ambivalent effects of added layers on steady kinematic dynamos
 in cylindrical geometry: application to the VKS experiment.
 \textit{Eur. J. Mech. B/Fluids},
2006c, \textbf{25}, 894-908.
\bibitem{TAKA}Takahashi, F., Matsushima, M., and Honkura, Y.,
Simulations of a quasi-Taylor state geomagnetic field including
polarity reversals on the Earth Simulator.
\textit{Science}, 2005, \textbf{309}, 459-461.
\bibitem{TARDUNO} Tarduno, J. A., Cottrell, R. D. and Smirnov, A. V.,
High geomagnetic intensity during the mid-Cretaceous from Thellier analysis of
simple plagioclase crystals. \textit{Science}, 2001, \textbf{291}, 1779.
\bibitem{VALET1993}Valet, J.-P. and Meynadier, L., Geomagnetic field intensities
and reversals during the last 4 million years. \textit{Nature}, 1993, \textbf{366},
 234-238
\bibitem{VALET2005}Valet, J.-P., Meynadier, L. and Guyodo, Y.,
Geomagnetic dipole strength and reversal rate over the
past two million years. \textit{Nature}, 2005, \textbf{435},
 802-805
\bibitem{vanderPol}van der Pol, B., On relaxation oscillations. \textit{Phil. Mag.},
1926, \textbf{2}, 978-992.
\bibitem{Weisshaar}Weisshaar, E, A numerical study of $\alpha^{2}$-dynamos with anisotropic
$\alpha$-effect. \textit{Geophys. Astrophys. Fluid Dyn.}, 1982,
\textbf{21}, 285-301.
\bibitem{Wicht}Wicht, J. and Olson, P., A detailed study of the polarity reversal
mechanism in a numerical dynamo model. \textit{Geochem. Geophys. Geosys.},
2004, \textbf{5}, Art. No Q03H10.
\bibitem{Xu-a}Xu, M., Stefani, F. and Gerbeth, G., The integral equation method
for a steady kinematic dynamo problem. \textit{J. Comp. Phys.}, 2004a,
\textbf{196}, 102-125. 
\bibitem{Xu-b}Xu, M., Stefani, F. and Gerbeth, G., Integral equation approach to
time-dependent kinematic dynamos in finite domains. \textit{Phys.
Rev. E}, 2004b, \textbf{70}, Art. No. 056305.
\bibitem{Yoshimura}Yoshimura, H., Wang, Z. and Wu, F., Linear astrophysical dynamos in
rotating spheres: mode transition between steady and oscillatory dynamos
as a function of dynamo strength and anisotropic turbulent diffusivity.
\textit{Astrophys. J.}, 1984, \textbf{283}, 870-878.
\end{thebibliography}
\end{document}